# Evidence for a Purely Double-Exchange Mechanism for the Anisotropic Conductivities in Layered La-Sr-Mn-O Single Crystals Below $T_C$

Qing'An Li, K.E. Gray, J.F. Mitchell, A. Berger and R. Osgood

*Materials Sciences Division, Argonne National Laboratory, Argonne, IL, 60439*

**PACS #**

## ABSTRACT

Experimental evidence supports the double-exchange (DE) mechanism for both in-plane and c-axis conductivity in the colossal-magnetoresistive (CMR) layered manganite $La_{1.4}Sr_{1.6}Mn_2O_7$. Below $T_C$, the data determine both the DE and antiferromagnetic (AF) superexchange between bilayers. These agree with recent spin-wave data, such that the ratio of intra- to inter-bilayer DE constants is very close to the spin-independent conductance anisotropy of ~300. The conductivity is shown to be proportional to the square of the measured magnetization over a wide range of fields and for temperatures below the in-plane ferromagnetic ordering temperature, $T_C$. This dependence is shown to be consistent with DE coupling and earlier zero-field studies of $La_{1.2}Sr_{1.8}Mn_2O_7$ near $T_C$, which used neutron scattering for the local magnetization. A mixed AF and spin-flop state (similar to the intermediate state of type-I superconductors) is a rigorous prediction of the data and modeling.

**Introduction.**  One of the most fundamental needs for understanding colossal magnetoresistive (CMR) oxides[1] is the connection of resistivity, $\rho$, or conductivity, $\sigma$, to magnetization, M, which can help address the nature of conduction, e.g., if it is by the double-exchange (DE) mechanism.  The layered CMR manganites[2] uniquely offer a wide range of opportunities to impact our understanding of the phenomena, since, (1) there is an extremely rich variety of magnetic structures reported for the bilayer compounds, $La_{2-2x}Sr_{1+2x}Mn_2O_7$, as a function of x, the hole doping, and (2) the anisotropic conductivity and magnetic order can be measured and affected independently.  In $La_{1.4}Sr_{1.6}Mn_2O_7$, hereafter referred to by its hole doping of 30%, ferromagnetic (FM) order is exhibited *within* the bilayers, but these are ordered antiferromagnetically (AF) along the c-axis[3].  The present study of this 30% composition, over a wide range of magnetic fields and temperatures, supports the DE mechanism for both in-plane and c-axis conductivity below the in-plane FM ordering temperature, $T_C$.  The data determine both the AF superexchange and the DE constants between adjacent bilayers.  These are in excellent agreement with recent spin-wave data[4], such that the ratio of intra- to inter-bilayer DE constants is very close to the spin-independent (i.e., parallel spin) conductivity anisotropy of ~300.  The conductivity is shown to be proportional to the square of the measured magnetization, $\sigma \sim M^2$, over a wide range of fields for temperatures below $T_C$.  This dependence is further shown to be consistent with DE coupling as well as with earlier zero-field studies[5] of $La_{1.2}Sr_{1.8}Mn_2O_7$ near $T_C$, which used neutron scattering to measure the local magnetization.  An interesting mixed AF and spin-flop state (similar to the intermediate state of type-I superconductors) is a rigorous prediction of the data and modeling.

While the layered manganites offer unique opportunities, it is not without a cost.  Single crystals must exhibit extremely small compositional variations and mosiac spread[6].  Multiterminal conductance measurements are required, with well-defined sample shapes and contact locations[7].  Beyond the competition between AF and DE coupling, higher harmonics of the superexchange and crystalline anisotropy are needed and the layering introduces additional dipole terms in the magnetic energies. Although these make the analysis somewhat tedious, the experimental data *overconstrains* the available parameters in such a way that the solution can be relied upon as complete and correct.

**Sample Fabrication and Characterization.** Crystals of the 30% material were melt-grown in flowing 20% $O_2$ (balance Ar) using a floating-zone optical image furnace (NEC- Model SC-M15HD). Samples for the measurements described in this work were cleaved from the resulting polycrystalline boules[6].

The relative widths of zero-field transitions are likely one of the best indicators of crystal quality. It is hard to imagine that any imhomogeneity or other inferior aspect of a sample's quality would sharpen the transition. The samples reported here exhibit the sharpest reported transition widths and minimal evidence of inhomogeneities, using tests described in Ref. 7.

**Transport Measurements.** Samples were measured in a standard $^4$He gas flow cryostat and heating was avoided by using small currents. For electrical contacts, Au was sputtered onto the c-axis-normal surfaces in a pattern of stripes perpendicular to the long dimension of the crystal. Four stripes were placed symmetrically on both the top and bottom surfaces. In the standard six-terminal configuration, the outer stripes on the top surface supply the current. The measured voltages between the inner stripes of the top and bottom surfaces are transformed into $\sigma_{ab}$ and $\sigma_c$ using the method of Ref. 7. The calculated $\sigma_{ab}$ and $\sigma_c$ for a 30% crystal are shown in Fig. 1, while the more customary resistivities are just the inverses. One new feature of the data is the large anisotropy which surpasses 10,000 below 30 K and is consistent with a c-axis AF ground state[3]. In fields of 7 T (either along the c-axis or in the ab-planes) the AF coupling is overcome and at temperatures below $T_C$, $\sigma_{ab}$ and $\sigma_c$ are shown in Fig. 1 to display the same shape and thus can be used to define a spin-independent anisotropy, which is

~300. This paper is mainly concerned with the dependences of $\sigma_{ab}$ and $\sigma_c$ on field for $T<T_C$ and fields up to saturation.

**Results.** Examples of $\sigma_{ab}$ and $\sigma_c$ at low temperatures (2-5 K) and 50 K are shown in Fig. 2a and its inset for applied fields, H, parallel to the ab-planes and in Fig. 2b for fields parallel to the c-axis. Note that $\sigma_{ab}$ and $\sigma_c$ are scaled by the spin-independent anisotropy to emphasize the virtually identical high-field behavior found above the saturation fields, e.g., $H^c_{s\,at} \sim$ 5,500 gauss for H||c, and up to the experimental maximum of 7 T. Above saturation, the AF c-axis correlations are completely overcome, but neither conductivity has reached full saturation, indicating that the magnetization, $M_s$, is less than the theoretical lattice sum, $M_{so}$. Further increases in $\sigma$ with H might result from reducing thermal fluctuations of Mn ions which are random and thus uncorrelated with conductivity direction.

For this composition, the crystalline anisotropy tends to align the individual moments along the c-axis at low temperatures[3] with AF coupling between fully FM bilayers. Then for a sufficiently large c-axis applied field, $H^c$, the Zeeman energy will overcome the net AF interbilayer exchange coupling. The $\sigma_c$ data in Fig. 2b show such an onset at ~1,100 gauss, however, the gradual increase in $\sigma_c$ between 1,100 and 5,500 gauss still requires an explanation. The large value of the demagnetization coefficient for these plate-like crystals (a=2.2 mm, b=0.3 mm and c=0.1 mm) can favor a mixture of AF and FM regions, akin to the intermediate state of type-I superconductors. Also, the lowest energy state in the FM regions may be a spin-flop (SF) state exhibiting finite angles $\pm\theta$ with respect to the c-axis which alternate in adjacent bilayers. The $\sigma_c$ data in Fig. 2b are qualitatively consistent with a mixed FM/AF or SF/AF state. Further support for the mixed state is found in the dip in $\sigma_{ab}$, seen at ~1,100 Oe in Fig. 2b. This is expected for a mixed state, since half of the FM coupled bilayers will develop AF orientations at the points separating the AF and SF states, thus reducing $\sigma_{ab}$ (see inset of Fig. 4 below).

**Analysis.** One expects particularly simple behavior for H∥ab as each of the two sublattices (i.e., alternating bilayers, one with c-axis spins up and the other with spins down) smoothly rotate towards the ab-plane without spin-flop or spin-flip transitions. Such cases are commonly found, and the low-field M is proportional to H. However, the data of Fig. 2a appear to be quadratic (plus a constant) for small H, implying that $\sim M^2$. The considerably smaller constant at 4.2 K suggests its origin is fluctuations[7]. To test whether $\sim M^2$, the normalized $_c$ is shown in Fig. 3a with $M/M_s$, measured on the same crystal. There is outstanding agreement except at the lowest fields. Although the large conductivity anisotropy invalidates the data below ~1000 Oe at low temperatures, $_c$ extrapolates to a finite intercept. This is in contrast to M(H) which goes through the origin with negligible hysteresis at all temperatures. These minor differences are readily understood by noting that $_c$ is related to the *local* magnetic order, whereas domain structures will null the net magnetization. A local FM component can result from a competition between DE and AF superexchange as well as spin-wave fluctuations[7]. The common absence of hysteresis in the layered manganites implies weak domain-wall pinning. Since the anisotropy is small at 50 K, $_c$ is accurately determined at all fields so the much larger differences in the inset of Fig. 3a are real and consistent with greater fluctuations.

At low temperatures, the finite intercept of $_c$ is understood (below) as a necessary result of finite DE while the deviation from linearity before saturation is consistent with higher order terms in the uniaxial crystalline anisotropy and/or in the AF interbilayer superexchange. These higher order terms are included in the energy per unit volume at the outset as $k'_u \sin^4$ and $J'_s \cos^2$ , where is the angle *between* spins in adjacent bilayers and is measured from the c-axis. For H∥ab, we assume the sublattice magnetizations are at and - , measured from the c-axis, so that = -2 . Minimizing the energy results in a cubic equation in $y = \sin$ :

$$y^3 + \left(\frac{H^a_{sat} M_s}{4G} - 1\right) y - \frac{J_d + H^a M_s}{4G} = 0 \tag{1}$$

where DE is introduced[8,9] as $J_d \cos(\phi/2)$, $G \equiv k'_u + 4J'_s$, $H^a_{s\ at}$ and $M_s$ are the saturation field and magnetization (which may be less than the theoretical lattice sum, $M_{so} \sim 480$ gauss). A fit of the low temperature data to Eq. 1 is shown in Fig. 3a for $H^a_{s\ at} = 15,100$ Oe and with $M_s = 480$ Oe it yields $J_d = 0.29 \times 10^6$ ergs/cm$^3$ and $G = 0.68 \times 10^6$ ergs/cm$^3$. Thus the data of Fig. 3a are well explained by $J_d$ and the higher order terms $k'_u$ and $J'_s$. A small misalignment of $H^a$ could cause[10] the slight rounding of the data near $H^a_{s\ at}$.

A definitve analysis at higher temperatures is hampered by the appearance of fluctuations, which at small fields (inset of Fig. 3a) reduce the nearly AF ground state spin alignment and thus produce a local moment. Here $\sigma_c(H=0)/\sigma_c^{s\ at}$ is approximately linear in temperature its extrapolates to T=0 yields the same $J_d$ as found from Fig. 3a. However, the above analysis cannot uniquely determine $J_d$ and the separate fluctuation contribution at higher temperatures. Above $H^a_{s\ at}$, fluctuations are *away* from the field-aligned paramagnetic state and the conductivity is very well described by $\sigma_c^{s\ at}\{1-\exp(-\mu H_{ab}/k_B T)\}$ with $\mu \sim 16 \mu_B$ ($\mu_B$=Bohr magneton).

For H‖c, the sublattice magnetizations after a SF transition will be at $\pm\theta$, so that $\phi = 2\theta$, and $y=\cos\theta$, but otherwise only the superscripts in Eq. 1 change from a to c. The fit, shown in Fig. 3b for $H^c_{s\ at}=6,000$ Oe and *the same* values of G and $J_d$, indicates that the data is only consistent with a uniform SF state very near to saturation, i.e., for $\cos\theta > 0.9$, and the significant disagreement at lower fields hints that a mixed AF/SF state is needed. Since the fit in Fig. 3b is constrained to be a SF state with neighboring bilayers at $\pm\theta$, the actual AF state at $H^c=0$ (see the $H^a=0$ limit of Fig. 3a) is not reproduced. Even smaller rounding of the data near $H^c_{s\ at}$ is seen, likely because field alignment is less crucial along the easy magnetic axis[10].

Having established that $\sigma \sim M^2$, the mixed AF/SF state is further and strongly supported by comparing the data in Fig. 3b to the measured magnetization, M, for H∥c. Three separate crystals, including the one used for transport, exhibit indistinguishable M(H) curves. Shown in Fig. 4 for 4.2 K are $M/M_s$ together with $\sigma_c$, scaled for the best fit below 4 kOe, and $\sigma_c$, normalized to its saturation value. The upper part of $M/M_s$ is best fit by the normalized $\sigma_c$ whereas the lower portion is only consistent with a scaled $\sigma_c$, and the crossover is fairly rigidly determined to be close to $\cos\theta \sim 0.8$-$0.9$ as found in Fig. 3b. If the lower portion describes a mixed AF/SF phase, then the spatially-averaged conductivity, $\sigma_c$, *and* magnetization, $M$, are both proportional to the *fraction*, f, of the sample in the SF state, and thus one expects $\sigma_c \sim M$, in good agreement with the data. Once the entire sample reaches the *uniform* SF state (f=1 and $\cos\theta \sim 0.87$), further increases in $H^c$ just rotate the spins closer to the c-axis. Thus the excellent qualitative comparisons of M to $\sigma_c$ seen in Fig. 4 strongly supports a mixed AF/SF state.

This picture should and can be made much more quantitative. In doing so, new issues as well as insights into the CMR state will emerge. While this model explains the overall shapes of the data in Figs. 2 and 3 in terms of an AF/SF mixed state, it is insufficient to correctly predict the onset at $H_1^c \sim 1,100$ Oe, as it stands. The internal dipole sum[11], which is nonzero in these layered structures, is needed. The regions between Mn-O bilayers are nonmagnetic and this dictates that the internal dipolar energies associated with the components of M perpendicular to the layers (i.e., along the c-axis) must be included[11]. Finally,

$$E_{SF}(H^c, \theta) = k_u \sin^2\theta + k_u' \sin^4\theta + J_s \cos\theta + J_s' \cos^2\theta - J_d \cos(\theta/2) - H^c M_s \cos\theta$$
$$+ 2\pi N_c M_s^2 \cos^2\theta + 2\pi D_f M_s^2 \cos^2\theta, \qquad (2)$$

$$E(H^a, \theta) = k_u \sin^2\theta + k_u' \sin^4\theta + J_s \cos\theta + J_s' \cos^2\theta - J_d \cos(\theta/2) - H^a M_s \sin\theta$$
$$+ 2\pi N_b M_s^2 \sin^2\theta + 2\pi D_a M_s^2 \cos^2\theta, \qquad (3)$$

where $J_s$ and $J_d$ are the absolute values of the interbilayer AF superexchange and FM double exchange energies per unit volume, $N_c$ and $N_b$ are the crystal's demagnetization coefficients for the experimental field directions, $D_f$ and $D_a$ are the calculated[11] dipolar energies for the pure FM and AF states and we assume that the sublattice magnetizations in Eq. 2 are at $\pm\phi$, measured from the c-axis, and are at $\phi$ and $\pi-\phi$ for Eq. 3. Then $\theta=2\phi$ and $\theta=\pi-2\phi$ for Eqns. 2 and 3, which become simply:

$$E(H^c, \phi) = G \cos^4\phi + L_c \cos^2\phi - (J_d + H^c M_s)\cos\phi - J'_s - J_s + K, \qquad (4)$$

$$E(H^a, \phi) = G \sin^4\phi + L_a \sin^2\phi - (J_d + H^a M_s)\sin\phi + J'_s - J_s + 2\pi D_a M_s^2, \qquad (5)$$

with $G \equiv k'_u + 4J'_s$, $K \equiv k_u + k'_u$, $L_c \equiv 2J_s + 2\pi(N_c + D_f)M_s^2 - G - K$ and $L_a \equiv 2J_s + 2\pi(N_b - D_a)M_s^2 - G + K$. The pure AF state is found from Eq. 5 with $\phi=0$, so that $E_{AF} = J'_s - J_s + 2\pi D_a M_s^2$.

Minimizing Eqns. 4 and 5 against $\phi$ results in similar cubic equations:

$$4G y^3 + 2L_x y - (J_d + H^x M_s) = 0 \qquad (6)$$

where $y = \cos\phi$ for $H \parallel c$ and $\sin\phi$ for $H \parallel ab$. Note that the double exchange term looks formally like an effective applied field. Then a pure AF state is seen to be unstable for finite $J_d$, even in zero field, against formation of a canted state[9]. To see this, take the low-field limit of Eq. 6 for x=a. Then $y \ll 1$, so $y^3$ can be neglected and $\sin\phi_0 \sim J_d/2L_a$, for which the lowest energy state has alternating bilayers at angles of $\phi_0$ and $\pi-\phi_0$. Note that $\phi_0$ gives a local ab-plane magnetization, but for $H^a=0$, domains will null the net magnetization. However, since $\rho_c$ only depends on *local* spin correlations, the effect of $\phi_0$ remains and is seen in Fig. 3a as the extrapolation of $\rho_c$ to a finite value in zero field. Note this effect of *local* spin correlations has already been seen in $\rho_{ab}$ of the fully 3D FM ordered 40% layered manganite[7].

For y=1 in Eq. 6, one has the following useful identifications:

$$L_x = \{J_d + H^x_{sat} M_s - 4G\}/2, \qquad (7)$$

which for x=a, yields

$$G = \{J_d + (H^a_{s\,at} - H_s)M_s\}/4, \tag{8}$$

and which are also useful to derive $J_s$ and K, as

$$J_s = \{J_d + (H^c_{s\,at} + H_s)M_s - 4(N_c + N_b + D_f - D_a)M_s^2\}/8, \tag{9}$$

$$K = \{(H^a_{s\,at} - H^c_{s\,at})M_s + 4(N_c - N_b + D_f + D_a)M_s^2\}/4. \tag{10}$$

Thus of the five unknown parameters of the model ($k_u$, $k'_u$, $J_s$, $J'_s$, and $J_d$), four independent combinations (G, K, $J_s$, and $J_d$) have been derived.

The free energy for the mixed AF/SF state, $E_M$, is now evaluated. If a fraction, f, of the AF state has transformed to the SF state, then the total energy is:

$$E_M = f\,E_{SF}(H^c, ) + (1-f)\,E_{AF}(H^c, _0) + 2\,N_c\bar{M}^2 + W\,f(1-f) \tag{11}$$

where $E_{SF}(H^c, )$ and $E_{AF}(H^c, _0)$ are found from the modifications of Eqns. 4 and 5 which follow, the spatially-averaged magnetization is $\bar{M} = f\,M_s\cos$ and W represents the combined effects of the AF/SF wall energy, due to *in-plane* FM double exchange, and the *local* field distortions of the nonuniform mixed state. Following the calculation of the intermediate state of type-I superconductors[12], estimates of the *maximum* value of W, using spin-wave data for the in-plane FM double exchange, give $\sim 0.04 \times 10^6$ ergs/cm$^3$, so the final term in Eq. 11 cannot be more than $\sim 0.01 \times 10^6$ ergs/cm$^3$. This is already small compared to other terms in the energy, but it can also be considerably reduced by spatially extended domain walls (similar to Bloch walls in conventional FM), so initially, W will be ignored. In the mixed state, the demagnetization energy is contained in the third term of Eq. 11, so $2\,N_c M_s^2$ must be subtracted from $L_c$ of Eq. 4 and note that it is avoided entirely in the initial SF region (f<<1), e.g., by forming a narrow column or laminar sheet parallel to the c-axis field. To evaluate $E_{AF}(H^c, \theta_0)$ from Eq. 5, note that the Zeeman energy cancels in alternating layers for a c-axis field, and domains null the net magnetization (since $H^a=0$), so $2\pi N_b M_s^2$ is subtracted from $L_a$. Thus

$$E_{SF}(H^c, \theta) = G\cos^4\theta + (L_c - 2\pi N_c M_s^2)\cos^2\theta - (J_d + H^c M_s)\cos\theta - J'_s - J_s + K, \quad (12)$$

$$E_{AF}(H^c, \theta_0) = G\sin^4\theta_0 + (L_a - 2\pi N_b M_s^2)\sin^2\theta_0 - J_d\sin\theta_0 + J'_s - J_s + 2\pi D_a M_s^2, \quad (13)$$

where $\sin\theta_0$ equals the extrapolated intercept of Fig. 3a, $y(0,0)=0.06$ (at 2 K to avoid fluctuation effects), so that $\theta_0 \sim 4°$. The leading correction to the pure AF state is $\sin^2\theta_0$, since $J_d = y(0,0)H_s M_s$, and using the calculated values[11] of $D_a=1.98$ and $D_f=1.04$, the energy of the canted zero-field 'AF' state is lower than the pure AF state energy by $\sim 0.009 \times 10^6$ ergs/cm$^3$.

Using Eqns. 12 and 13, Eq. 11 is rewritten as

$$E_M = Af + Bf^2 + E_{AF}(H^c, \theta_0)$$

$$A = Gy^4 + (L_c - 2\pi N_c M_s^2)y^2 - (J_d + H^c M_s)y - 2\pi D_a M_s^2 - 2J'_s + K + W$$

$$B = 2\pi N_c M_s^2 y^2 - W \quad (14)$$

which, upon minimization with respect to f, yields

$$f_{min} = -A/2B \quad (15)$$

$$E_M(f_{min}) = -A^2/4B + E_{AF}(H^c, \theta_0) \quad (16)$$

where $E_M(f_{min})$ must be minimized with respect to $y=\cos\theta$. If W can be neglected, this can be solved analytically from the quadratic equation in $u=y^2$:

$$3Gu^2 + (L_c - 2\pi N_c M_s^2)u - K + 2\pi D_a M_s^2 + 2J'_s = 0, \quad (17)$$

which gives $u_{min}$ or $y_{min}$ or $\phi_{min}$ which are *independent* of applied field, $H^c$. Then the normalized magnetization $M/M_s$ (or equivalently $\sqrt{\chi_c/\chi_{cs}}$ ) is just

$$M/M_s = f_{min}(H^c) y_{min} = -A y_{min}/2B. \qquad (18)$$

From this it is easy to show that the slope of $M/M_s$ versus $H^c$ is just $1/4\pi N_c M_s$. This is equivalent to using the internal field, $H_{int}$, as given by an ellipsoid, i.e., $-4\pi N_c M_s$. From the initial slope of Fig. 4, one finds $4\pi N_c M_s \sim 3510$ gauss. Using $N_c \sim 0.78$, found by approximating our rectangular crystal by an ellipsoid, one gets $M_s \sim 360$ gauss. This is smaller than the lattice sum, $M_{so}=480$ gauss, which is in excellent agreement with the measured $M_s$ from two magnetometers on two different crystals[13]. Our rectangular parallelpiped crystal is likely better described by a smaller $N_c$ (~0.58).

There is one free parameter, e.g., $J'_s$, which is not yet determined by $H^a_{s\,at}$, $H^c_{s\,at}$, $H_s$ and $y(0)$. Note that the experimental results *dictate several requirements to be satisfied by the one remaining free parameter*: (1) that $f=0$ for $H^c=H^c_1 \sim 1200$ Oe; (2) that $f=1$ for $H^c=H^c_2$ ~4200 Oe using the value of y found by minimizing the total energy of Eq. 11 with respect to $\phi$; (3) that y must be ~0.87; and finally, (4) the energy of the SF state must fall below that of the AF state at $H^c_1$. Although these requirements are not independent, it is an overconstrained situation such that if agreement with experiment can be found, it will lend strong support to the *completeness and correctness* of the model. This is the case: using $N_c=0.58$, $M_s=M_{so}=480$ Oe and $W=0$, the above equations determine the energy crossover ($f\sim 0$) at $H^c_1 \sim 1180$ Oe providing $J'_s = -0.07 \times 10^6$ ergs/cm$^3$. Note that $J'_s$ just modifies the angular dependence of the superexchange, so a negative value is acceptable if, as is the case here, $J'_s << J_s$ to maintain the AF character. The solution, shown as the solid line in Fig. 4, also gives $y=0.89$ ($\phi=26°$) throughout the mixed state, $f=1$ at $H^c \sim 4100$ Oe, $k_u=2.6 \times 10^6$ ergs/cm$^3$, $k'_u=0.96 \times 10^6$ ergs/cm$^3$, $J_s=1.0 \times 10^6$ ergs/cm$^3$ and $J_d=0.29 \times 10^6$ ergs/cm$^3$.

The veracity of these numerical values should not be overestimated. Each is proportional to the value of $M_s$ used, and in some cases a smaller part going as $M_s^2$, while $J_s'$ and $k_u'$ are especially sensitive to experimental accuracy. Including the largest potential value of W (0.04x10$^6$ ergs/cm$^3$) makes only small differences to the parameter values, but the solution no longer has precisely constant θ in the mixed state, and actually increases a few degrees from $H_1^c$ (f=0) to $H_2^c$ (f=1), followed by a decrease to zero at saturation. However, the mixed-state magnetization, $M/M_s$=yf, for finite W is still strictly linear in $H^c$. Of greatest importance is the consistent understanding of a large body of experimental data using the terms in Eqns. 2 and 3 with sensible values for the parameters. As a result the data and analysis strongly support the contention of a mixed SF/AF state in this material.

Although the overall fit is tightly constrained, it is of interest to ascertain how rigorously various of the constants are determined. Of particular interest is $J_d$, which comes from the extrapolation of ρ$_c$ vs. $H^a$ as shown in Fig. 3a. If $J_d$ is artificially set equal to zero in the above procedure for H‖c, acceptable agreement can be found with the most notable differences being $J_s'$, $k_u'$ and θ. Although this agreement with experiment for H‖c does not add direct further justification for DE, it implies the completeness and correctness of the model used with the data of Fig. 3a to determine the existence, and magnitude, of the DE constant, $J_d$.

For the DE mechanism to describe the anisotropic ρ, it is necessary to establish consistency with Δρ~$M^2$. Using the square of the DE transfer matrix element[14], $t_o$cos(θ/2), one can easily show that for neighboring spins which are *uncorrelated* (even if on average they yield a net magnetization, M, meaning that each is individually correlated with a magnetization axis), that ρ~1+$(M/M_s)^2$. This situation, which might occur for T>>$T_C$, is discussed in Ref. 14, together with an explanation of the small Δρ found above $T_C$. However, this calculation is changed for strong FM nearest-neighbor correlations (e.g., as expected for DE below $T_C$). The applied field favors the predominant alignment of the pair magnetizations along its axis, which for the cases considered above (e.g., a SF state consisting of neighboring bilayers with spins at ± with respect to the c-axis) result in $M \sim M_s \cos(\theta/2)$ and thus $\sigma \sim (M/M_s)^2$ for DE.

The *intra*layer exchange coupling, $J_1$, assumed to be DE, can be estimated from spin-wave data[4] for the 40% sample to be ~8.5 meV/bond. It is likely close to the same value for the 30% crystal, so $J_1$ is ~300 times larger than the *inter*bilayer $J_3 \sim 0.027$ meV/bond found from the above value of $J_d$. The similarity to the spin-independent conductance anisotropy of ~300 is surprising, since in the DE model, the *conductance* per bond is $\sim J^2 D$, where $D$ is the density of final states. Note that for *conductivity*, geometrical considerations in the layered manganites reduce the ratio $J_1^2/J_3^2$ by $(s/a)^2/8 \sim 3$, so that one would conclude that $D_c/D_{ab} \sim 100$.

**Discussion.** The two main results of this research are (1) the identification of DE from the low-temperature $\sigma_c$ extrapolated to zero field and (2) the demonstration that $\sigma \sim M^2$ over a wide range of fields for $T<T_C$. It was also shown that (2) is the expectation of DE and therefore adds further support to the identification of (1). However, earlier work[15] on thin films of the 3D perovskite CMR, $La_{0.7}Ca_{0.3}MnO_3$, concluded a significantly different connection than (2). They found $\sigma \sim \exp(M/M_0)$ and ascribed it to polaronic hopping transport within a DE framework. Since our data spans the full range of M up to $M_s$, it is unlikely that such an exponential dependence was missed. Therefore, other reasons must be found for the difference. One possibility is the suggestion of Furukawa[16] that the La-Ca-Mn-O perovskite, unlike La-Sr-Mn-O, cannot be understood in a simple DE framework.

**Summary.** In addition to the identification of the DE mechanism and the demonstration that ~$M^2$, we have presented credible evidence for a mixed spin-flop/antiferromagnetic state similar to the intermediate state of type-I superconductors. Evidence is presented supporting significant thermal fluctuations *away* from both the AF ground state at low fields and the field-aligned paramagnetic state. Detailed analysis reveals various parameters specific to the 30% composition. These include the crystalline anisotropy, AF superexchange and the double exchange. Comparing the spin-independent conductivity anisotropy with the DE anisotropy, one concludes the density of states is 100 times larger for c-axis transport.

**Acknowledgements.** The authors acknowledge very useful ongoing discussions with Marcos Grimsditch, Ray Osborn, Stefan Rosenkranz, Sam Bader and Denis Golosov and the results of magnetization measurements by Ulrich Welp. This research is supported by the U.S. Department of Energy, Basic Energy Sciences-Materials Sciences, under contract #W-31-109-ENG-38.

# REFERENCES

1. S. Jin, T.H. Tiefel, M. McCormack, R.A. Fastnacht, R. Ramesh and L.H. Chen, *Science* **264**, 413 (1994).

2. Y. Moritomo, A. Asamitsu, H. Kuwahara and Y. Tokura, *Nature* **380**, 141 (1996).

3. T. Kimura, Y. Tomioka, H. Kuwahara, A. Asamitsu, M. Tamura and Y. Tokura, *Science*, **247**, 1698 (1996).

4. S. Rosenkranz, R. Osborn, L. Vasiliu-Doloc, J.W. Lynn, S.K. Sinha and J.F. Mitchell, private communication.

5. R. Osborn, S. Rosenkranz, D.N. Argyiou, L. Vasiliu-Doloc, J.W. Lynn, S.K. Sinha, J.F. Mitchell, K.E. Gray and S.D. Bader, *Phys. Rev. Lett.* **81**, 3964 (1998).

6. J.F. Mitchell, D.N. Argyriou, J.D. Jorgensen, D.G. Hinks, C.D. Potter and S.D. Bader, *Phys. Rev.* **B55**, 63 (1997).

7. Qing'An Li, K.E. Gray and J.F. Mitchell, *Phys. Rev.* **B**. (accepted).

8. C. Zener, *Phys. Rev.* **82**, 403 (1951) and P.W. Anderson and H. Hasegawa, *Phys. Rev.* **100**, 675 (1955).

9. P.G. de Gennes, *Phys. Rev.* **118**, 141 (1960).

10. M. Grimsditch, E.E. Fullerton and R.L. Stamps, *Phys. Rev.* **B56**, 2617 (1997).

11. A. Berger, private communication.

12. M. Tinkham, *Introduction to Superconductivity*, (McGraw-Hill, New York, 1975).

13. U. Welp, private communication.

14. A.J. Millis, P.B. Littlewood and B.I. Shraiman, *Phys. Rev. Lett.* **74**, 5144 (1995).

15. M.F. Hundley, M. Hawley, R.H. Heffner, Q.X. Jia, J.J. Neumeier, J. Tesmer, J.D. Thompson and X.D. Xu, *Appl. Phys. Lett.* **67**, 860 (1995).

16. N. Furukawa, private communication.

# Figure Captions

Fig. 1.   The principle anisotropic conductivities, $\sigma_{ab}$ (circles) and $\sigma_c$ (diamonds), as derived from data on a 30% crystal.  Note the insulating behavior of $\sigma_c$ below 60 K.  Data in applied fields of 7 T, either parallel to the c-axis or ab-plane, align the spins to define the spin-independent anisotropy of ~300.  The insets show the AF spin arrangement in zero field and the aligned states at 7 T.  Insets: spin configurations, where the arrows represent both spins in a bilayer, which are offset by half a lattice constant in neighboring bilayers.

Fig. 2.   (a) The field dependences (H∥ab) of $\sigma_{ab}$ (line) and $\sigma_c$ (squares) at 2 K, which have been scaled by the spin-independent anisotropy.  Inset: same for 50 K.  (b) Same as (a) except H∥c, and T=4.2 and 50 K.

Fig. 3.   (a) The field dependences (H∥ab) of $\sigma_c/\sigma_c^{sat}$ (solid line) and the measured magnetization, $M/M_{sat}$ (squares).  The dashed line (circles) is a fit to the data using the model described in the text.  Inset: same for 50 K but without fit.  (b) Same as (a) except H∥c and without $M/M_{sat}$.

Fig. 4.   The field dependences (H∥c) of $M/M_{sat}$ (diamonds), $\sigma_c/\sigma_c^{sat}$ (circles) and $\sigma_c$ (squares) which has been scaled for the best fit to M for H<4 kOe.  Clearly neither $\sigma_c$ or $\sigma_c/\sigma_c^{sat}$ fit M over the whole field range.  The solid line is a fit to the mixed AF/SF state model described in the text.  Inset: Schematic spin arrangement in the mixed state, in which the arrows represent both spins in a bilayer and, for simplicity, the finite angles, $\phi$ in the SF state and $\phi_o$ in the AF state, are ignored.